\documentstyle [12pt,preprint,eqsecnum,aps]{revtex}
\begin {document}
\tightenlines
\draft

\title {Phonon-Coupled Electron Tunneling in Two and Three-Dimensional
           Tunneling Configurations}

\author {E.Pazy and B.Laikhtman}
\address {Racah Institute of Physics, Hebrew University, Jerusalem 91904,
Israel}

\maketitle \begin{abstract}

We treat a tunneling electron coupled to acoustical phonons through a
realistic electron phonon interaction: deformation potential and
piezoelectric, in two or three-dimensional tunneling configurations. Making
use of slowness of the phonon system compared to electron tunneling, and
using a Green function method for imaginary time, we are able to calculate
the change in the transition probability due to the coupling to phonons. It
is shown using standard renormalization procedure that, contrary to the
one-dimensional case, second order perturbation theory is sufficient in order
to treat the deformation potential coupling, which leads to a small
correction to the transmission coefficient prefactor. In the case of
piezoelectric coupling, which is found to be closely related to the
piezoelectric polaron problem, vertex corrections need to be considered.
Summing leading logarithmic terms, we show that the piezoelectric coupling
leads to a significant change of the transmission coefficient.

\end{abstract}

\section {Introduction }
\label {sec:Introduction}

The effect of a bosonic environment on a tunneling particle has
been extensively studied. The work of Caldeira and
Leggett \cite{Caldeira 81}, which originally studied the tunneling
of a particle from a metastable state, focused attention on this
problem. Today there are many separate fields devoted to related
questions. As an example the case of coupling of a tunneling
particle to a photonic
environment is the basis of three such fields: that of dynamic
image potential \cite{Perrson88},
Coulomb blockade \cite{Grabert92} and dissipative macroscopic
quantum tunneling (MQT) of
Josephson phase \cite{Devoret92}.
The development of experimental techniques through
which different effects now become within the scope
of experimental verification \cite{Marclay88} is a
source of enhanced interest in such problems.

The coupling of a tunneling electron to phonons has received much less
attention \cite{Bak86}, though this problem is of fundamental interest, being
closely related to the polaron problem \cite{re:polaron}, as well as being of
potentially great technological importance. In a previous paper \cite{pazy99}
we have shown that in the one-dimensional case, i.e., a quantum wire or
narrow constriction, the coupling to the zero-point fluctuations of the
acoustic phonon field leads to an exponential enhancement of the tunneling
probability. The question of what happens to this enhancement in the the
two-dimensional case in which the electron's wave function is restricted in
one of the the directions perpendicular to the tunneling direction, and the
three-dimensional case where the electron is free to move in both the
perpendicular directions, is the subject of this paper. We address the
problem of an electron tunneling coupled to acoustic phonons via two
realistic coupling mechanisms: deformation potential and piezoelectric. For
both coupling mechanisms we calculated the change of the tunneling
probability and the renormalization of the potential barrier height. In two
and three dimensions the change of tunneling probability is smaller than in
one dimensional case and for its calculation we developed a perturbation
theory employing the Green function method. The problem of calculation of the
phonon correction, however, appeared to be nontrivial for the piezoelectric
coupling, where the summation of high order terms of the perturbation theory
is necessary.

Being interested in the effect of zero-point fluctuations of the phonon field
on the tunneling electron we did not consider the coupling to optical
phonons. The virtual process of phonon emission and absorption is inversely
proportional to the energy difference involved in that process. Since
emission and absorption of optical phonons involves a large energy difference
the interaction with optical phonons can be neglected.

To calculate the transmission coefficient we make use of an approximation
based on the slowness of the phonon field compared to the tunneling electron,
which we refer to as the static approximation \cite{pazy99}, and obtain a
stationary Schr\"{o}dinger equation for the electron coupled to phonons. We
then apply the parabolic (paraxial) approximation which is equivalent to
considering the following physical situation (which is typical
experimentally) (i) the barrier is so high that the tunneling can be
considered semi-classically and (ii) the interaction energy between electrons
and phonons is small compared to the height of the barrier. Treating the
phonons via a Green function method the self-energy term is then calculated
using second order perturbation theory. The resulting self-energy term is
divergent at large transferred momenta, for both coupling mechanisms, and
these divergences are eliminated by a renormalization procedure. For the
deformation potential one has to renormalize the potential barrier height and
the electron's effective mass, this finally leads to a correction to the
transition probability. On the other hand for the piezoelectric coupling,
after the renormalization of the potential barrier height the second order
correction diverges logarithmically. The summation of main logarithmic terms
resulted in a factor non-trivially depending on the electron-phonon coupling
constant and can lead to a significantly change the transmission coefficient.

The paper is composed in the following way. In Sec.\ref{sec:form} we
formulate the problem, introduce the model, and present the static
approximation which helps us facilitate the calculations. In
Sec.\ref{sec:green} the Green function method through which we approach the
problem is described, the parabolic approximation is introduced, self-energy
diagrams are presented and it is shown how one uses them to derive the
transition probability. In Sec.\ref{sec:deformation} the renormalized
self-energies for the two and three-dimensional tunneling configurations are
given for the deformation potential coupling and the renormalization
equations are presented. The resulting change in the transmission coefficient
is calculated. In Sec.\ref{sec:piezoelectric} the method for including vertex
corrections for the piezoelectric case is given and the resulting change in
the transmission coefficient is calculated. The calculation of some
complicated integrals is given in appendices.

\section{Formulation of the problem}
\label{sec:form}
\subsection{The model}
\label{sec:model}

We treat the problem of an electron coupled to acoustic phonons while
tunneling through a rectangular barrier, of width L. We will treat two cases
that of a two and three-dimensional tunneling configurations. In the
three-dimensional, case the electron is free to move in the perpendicular (to
the tunneling) directions, whereas in the two-dimensional case the electron's
movement is restricted in one of the perpendicular directions i.e a two
dimensional electron gas. In both cases phonons move freely through the bulk
and thus are treated as three-dimensional. Being interested in the effect of
the zero-point fluctuations of the phonon field on the electron tunneling
probability we consider the system at zero temperature.

\subsection{Static approximation}
\label{static}
In the general case of a tunneling particle coupled to the environment two
typical time scales arise: the traverse time for the tunneling particle and
the typical time in which the environment reacts to the tunneling of the
particle. When the tunneling particle is an electron and the environment
acoustical phonons, it can be shown that typically the traverse time for the
electron is much smaller than the environmental reaction time scale. The
small parameter expressing the small ratio between these time-scales is:
$s/v$ , where $s$ is the phonon sound velocity, and $v$ is the under the
barrier electron semi-classical velocity.\cite{pazy99}  For a potential
barrier of around 10 meV in bulk GaAs, this ratio is of the order of 0.01.
This small ratio leads us to what we refer to as the static approximation. In
this approximation the phonon dynamics is neglected and the potential created
by the barrier has a phonon addition that may have different time independent
shapes with the probability equal the wave function squared of the phonon
field. [For a more detailed description of the static approximation one is
referred to the previous paper \cite{pazy99}.] As a result the electron wave
function underneath the barrier satisfies the following stationary
Schr\"{o}dinger equation for the electron coupled to phonons

\begin{equation}
\label{eq:a1}
-\frac{{\hbar}^2}{2 m}{\nabla}^2 \Psi({\bf x},u_{\bf q})+ (V + f({\bf x},
u_{\bf q})) \Psi({\bf x},u_{q})=
E \Psi({\bf x},u_{\bf q}),
\end{equation}
where ${\bf x}$ is the electron coordinate, $m$ is the electrons effective
mass, $V$ is the potential barrier height, $u_{\bf q}={u_{\bf -q}}^*$ are the
Fourier coefficients of the lattice displacement vector $u({\bf x})$. $u_{\bf
q}$ is a random variable whose probability is controlled by the ground state
phonon wave function. For the two-dimensional case the Laplacian is the
two-dimensional Laplacian operator. $f({\bf x},u_{\bf q})$ is the interaction
term between the electron and phonons defined in the following way. For the
three-dimensional case, in which the electron wave function is not restricted
in the perpendicular to tunneling direction which is taken as $x$, the
deformation potential coupling interaction term is given by

\begin{equation}
f_{d}({\bf x}, u_{\bf q})=
\frac{i\Lambda}{\sqrt{{V}_{vol}}} \sum_{\bf q}  \vert {\bf q} \vert
u_{\bf q} \exp (i {\bf q x} )  \ ,
\label{eq:a2}
\end{equation}
where $\Lambda$ is the deformation potential constant and ${V}_{vol}$ is the
normalization volume. For the piezoelectric coupling the interaction term in
the Hamiltonian is given by
\begin{equation}
f_{p}({\bf x},u_{\bf q})=
\frac{\Xi}{\sqrt{{V}_{vol}}}
\sum_{\bf q} u_{\bf q}
\exp (i{\bf q x}) ,
\label{eq:a3}
\end{equation}
where $\Xi =4\pi
e\sqrt{\langle{{\beta_{ijk}}^{2}\rangle_{\Omega}}/\epsilon}$, $e$ is the
electron charge, $\epsilon$ is the dielectric constant, $\beta_{ijk}$ is the
piezoelectric modules and ${\langle ...\rangle}_{\Omega}$ denotes the average
over angles. Though the piezoelectric interaction is anisotropic the
isotropic Hamiltonian approximation is commonly used \cite{re:isotropic}.
Subscripts $d$ and $p$ shall be used
throughout this paper, unless otherwise stated to define deformation
potential and piezoelectric related terms respectively.

For two dimensional electron gas were the electron wave function is taken as
restricted in the $z$ direction the interaction terms are given by
\begin{equation}
f_{d}({\bf x}, u_{\bf q})=
\frac{i\Lambda}{\sqrt{{V}_{vol}}} \sum_{\bf q} \ \vert {\bf q} \vert \
u_{\bf q} \ M_{q_{z}} \ \exp{[i(q_{x} x + q_{y} y)]},
\label{eq:a5}
\end{equation}
for the deformation potential coupling and
\begin{equation}
f_{p}({\bf x},u_{\bf q})=
\frac{\Xi}{\sqrt{{V}_{vol}}}
\sum_{\bf q} u_{\bf q}\  M_{q_{z}}\ \exp {[i(q_{x} x + q_{y} y)]}.
\label{eq:a6}
\end{equation}
for the piezoelectric coupling. Here $M_{q_{z}}$ is the matrix element of the
phonon exponent, $\exp{(i q_{z} z)}$ between the wave functions describing
the electron quantization in the $z$ direction.

We neglect screening of the electron-phonon interaction, since underneath the
potential barrier there are no free electrons and screening by remote
electrons is small.

Although the static approximation resembles the electron tunneling in the
presence of static disorder there is a very important difference between the
two problems. Whereas a static random field can absorb any momentum from the
electron the momentum, absorption in the static approximation is related to
phonon emission. Since under the static approximation it is assumed that the
phonon configuration does not change the tunneling electron momentum is
conserved.

\section{Green function method}
\label{sec:green}

The commonly used approach for addressing the problem of a tunneling particle
coupled to the environment is based on the path-integral approach. The
path-integral approach to the problem is extremely convenient in the
one-dimensional case where the contribution to the transition probability is
essentially from one trajectory. In two and three dimensional cases many
different trajectories contribute to the transition probability and we prefer
to use Green function method. The application of this method is facilitated
by the fact that a reasonable potential barrier (e.g., 10 meV) it is much
larger than electron-phonon interaction, so for the calculation of electron
Green function we can use perturbation theory. In some points the Green
function method that we use differs from commonly used formalism and we
present it's derivation.

\subsection{Parabolic approximation}
\label{sec:parabolic}

We consider the case when the coupling between the electron and the acoustic
phonons is small compared to the effective potential barrier height $V-E$,
which is considered as large. More than this, we assume that
$\hbar/\sqrt{2m(V-E)}$ is the smallest length scale in the problem. It is
much smaller than the width of the barrier $L$ and in two-dimensional case it
is much smaller than the size of the quantization in $z$ direction, $d$. The
first of these conditions enables us to use the parabolic (paraxial)
approximation \cite{Babic91}, and is the basis for our perturbative method.

Under the parabolic approximation we search for a solution to the
Schr\"{o}dinger equation (\ref{eq:a1}) of the form: $\Psi({\bf x},u_{q})=\exp
(-\kappa x) \psi({\bf x},u_{q})$ where $x$ is the tunneling direction, the
exponential factor characterizes tunneling without interaction with phonons,
$\kappa=\sqrt{2m (V-E)}/\hbar$, and $\psi({\bf x},u_{q})$ is a slowly varying
envelope function. Slow variation of $\psi({\bf x},u_{q})$ follows from the
weakness of electron-phonon interaction and leads to the possibility of
neglect of its second derivative with respect to $x$,
$(\partial^2\psi/\partial{x}^2)\ll\kappa(\partial\psi/\partial x)$. Then Eq.
(\ref{eq:a1}) becomes
\begin{equation}
\label{eq:gfm.pa.1}
-\frac{1}{2}{{\nabla}^{2}}_{\perp}\psi + \kappa
\frac{\partial \psi}{\partial x} = - \frac{m}{{\hbar}^2}
f({\bf x}, u_{\bf q})\psi,
\end{equation}
where ${\nabla}_{\perp}$ designates the gradient with respect to the
perpendicular to tunneling direction $x$ (in the two-dimensional case
${\nabla}_{\perp}=\partial/\partial y$). This Schr\"{o}dinger equation is
just equivalent to a non-stationary Schr\"{o}dinger equation where $x/\kappa$
plays the role of imaginary time and there are two spatial dimensions instead
of the original three.

The transmission across the barrier is characterized by the solution to
Eq.(\ref{eq:gfm.pa.1}) with the boundary ("initial") condition $\psi(x=0,{\bf
r}_{\perp})=1$. It can be written as
\begin{equation}
\psi(x,{\bf r}_{\perp}|u_{\bf q}) = \int
    {\cal G}(x,{\bf r}_{\perp};0,{\bf r}_{\perp}^{\prime}|u_{\bf q})
    \psi(0,{\bf r}_{\perp}^{\prime}) d{\bf r}_{\perp}^{\prime} \ .
\label{eq:gfm.pa.2}
\end{equation}
Then the transmission coefficient is given by (multiplied by $\exp(- 2\kappa L)$
which is due to the parabolic approximation)
\begin{equation}
\left\langle\left|\psi(L,{\bf r}_{\perp}|u_{\bf q})\right|^{2}\right\rangle =
    \int    \left\langle
    {\cal G}(L,{\bf r}_{\perp};0,{\bf r}_{\perp}^{\prime}|u_{\bf q})
    {\cal G}^{\ast}(L,{\bf r}_{\perp};0,{\bf r}_{\perp}^{\prime\prime}|u_{\bf q})
            \right\rangle
    \psi(0,{\bf r}_{\perp}^{\prime}) \psi(0,{\bf r}_{\perp}^{\prime\prime})
    d{\bf r}_{\perp}^{\prime} d{\bf r}_{\perp}^{\prime\prime} \ ,
\label{eq:gfm.pa.3}
\end{equation}
where the angular brackets mean the averaging over the phonon configuration.
Due to translational invariance the average of the Green function product
depend on the difference ${\bf r}_{\perp}^{\prime}-{\bf
r}_{\perp}^{\prime\prime}$ so that the rhs does not depend of ${\bf
r}_{\perp}$.

In the case of tunneling in the presence of a random field the average of the
Green function product mixes both Green functions, i.e., it is not equal to
the product of averages. The situation is much simpler in the case of phonons
under the static approximation. The neglect of phonon emission means that
$\psi(L,{\bf r}_{\perp}|u_{\bf q})$ in the lhs of Eq.(\ref{eq:gfm.pa.3})
corresponds to the phonon ground state. In other words, for the calculation
of the transmission coefficient we have to keep only the diagonal with
respect to phonons part of the Green function. This diagonal part is obtained
by the averaging of $\cal G$ over the phonon configuration, $G(x,{\bf
r}_{\perp}-{\bf r}_{\perp}^{\prime})=\langle{\cal G}(x,{\bf r}_{\perp};0,{\bf
r}_{\perp}^{\prime}|u_{\bf q})\rangle$.

It is convenient to calculate $G(x,{\bf r}_{\perp}-{\bf r}_{\perp}^{\prime})$
in the Fourier representation. Actually we perform
Fourier transformation for the perpendicular direction and a Laplace
transformation for the $x$ direction. After these transformations
Eq.(\ref{eq:gfm.pa.1}) becomes
\begin{equation}
\label{eq:gfm.pa.4}
\frac{1}{2}k_{\perp}^2 \tilde{\psi}_{{\bf k_{\perp}}}(p) + \kappa
 \delta_{{\bf k_{\perp}},0} + \kappa p \tilde{\psi}_{{\bf k_{\perp}}}(p)
 =\sum_{\bf q_{\perp}}\sum_{q_{x}}F_{\bf q_{\perp}}(q_{x})
\tilde{\psi}_{{\bf k_{\perp}}-{\bf q_{\perp}}}(p-\imath q_{x}),
\end{equation}
where $\tilde{\psi}_{{\bf k_{\perp}}}(p)$ are the Fourier and Laplace
transforms of $\psi({\bf x},u_{q})$ with respect to the perpendicular and $x$
directions respectively and $F_{\bf q_{\perp}}(q_{x})=-(m/\hbar^2) f_{\bf
q_{\perp}}(q_{x})$. Throughout the paper we will use the notation $p$ for the
Laplace transformation parameter (imaginary $x$-component of the electron
wave vector), ${\bf k_{\perp}}$ for the electron wave vector in the plane
perpendicular to the tunneling direction and ${\bf q}$ for the phonon wave
vector.

The transmission coefficient now is $|t|^{2} e^{-2 \kappa L}$ where
\begin{equation}
\label{eq:gfm.pa.5}
t=\frac{1}{2\pi \imath} \int_{-\imath \infty +\tau_{0}}^{\imath \infty
+\tau_{0}} dp \ \ {e^{pL}} \ \ G_{k_{\perp}=0}(p) \ .
\end{equation}
Calculating the transmission coefficient thus amounts to calculating the
averaged Green function.

\subsection{Green function calculation}
\label{sec:diagram}

The Green function unperturbed by the phonon potential is
$G_{k_{\perp}}^{(0)}(p)=(\kappa p + k_{\perp}^2/2)^{-1}$ so
\begin{equation}
\label{eq:gfm.pa.6}
G_{k_{\perp}}(p)=
    \left[\kappa p + k_{\perp}^2/2 - \Sigma_{\bf k_{\perp}}(p)\right]^{-1} ,
\end{equation}
where $\Sigma_{\bf k_{\perp}}(p)$ is the self energy.

We calculate the self-energy with the help of a diagrammatic technique that
differs from the regular one in random field by two points. The first is that
the energy variable is replaced by $-\kappa p$. The other point, that in the
static approximation the two-particle Green function, Eq.(\ref{eq:gfm.pa.3}),
equals the product of two one-particle Green functions, has already been
mentioned. It is also possible to demonstrate this last point diagrammaticly,
showing that in the static approximation, due to momentum
conservation, diagrams with random field lines connecting two Green functions
contain the factor $1/V$, where $V$ is the volume of the system.

Due to the weak coupling between electrons and phonons, in most of the cases
that we consider a good approximation of the self-energy is given by the
first order diagram presented in Fig.\ref{fig1}, i.e.,
\begin{equation}
\label{eq:d1}
\Sigma_{\bf k_{\perp}}(p) = \sum_{{\bf q_{\perp}},q_{x}}
\langle {\vert F_{{\bf k_{\perp}}-{\bf q_{\perp}}}(q_{x})\vert}^2 \rangle
G_{{\bf q_{\perp}}}^{(0)}(p-\imath q_{x}),
\end{equation}

The averaging over phonons is carried out with the help of the phonon wave
function squared, $C\exp(-\sum_{\bf q}\rho|{\bf q}|s|u_{\bf q}|^2/\hbar)$,
where $\rho$ is the the crystal density $s$ is the sound velocity, and $C$ is
the normalization constant. Since $F_{{\bf q_{\perp}}}(q_{x})$ is
proportional to $u_{\bf q}$ the averaging gives $\langle F_{\bf q_{\perp}}
F_{\bf q'_{\perp}}
\rangle =\alpha {\vert q \vert}^{\pm 1} \delta_{\bf q_{\perp},
-q'_{\perp}}$. The $+1$ corresponding to the deformation potential coupling
and the $-1$ sign corresponding to the piezoelectric coupling. For the
deformation potential coupling $\alpha $ is a dimension-less parameter given
by

\begin{equation}
\label{eq:d2}
\alpha_{d}= \frac{\hbar /\rho s}{{\hbar}^4 / m^2}{\Lambda}^2 \ .
\end{equation}

For the piezoelectric coupling mechanism $\alpha$ has units of one over
length squared and is defined as

\begin{equation}
\label{eq:d3}
\alpha_{p}=\frac{\hbar /\rho s}{{\hbar}^4 / m^2} {\Xi}^2 \ .
\end{equation}

For the calculation of the transmission coefficient (\ref{eq:gfm.pa.5}) we
need the self-energy only for $k_{\perp}=0$,

\begin{equation}
\label{eq:d4}
\Sigma_{{\bf k_{\perp}}=0}(p)
= \frac{\alpha_{d,p}}{(2 \pi)^3} \int_{-\infty}^{\infty}
dq_{x} \int d^{2}{\bf q_{\perp}} \frac{1}{\kappa(p-\imath q_{x})+
\frac{1}{2}{\bf q_{\perp}}^{2}}{ \Bigl( \sqrt{{\bf q_{\perp}}^2 +{q_{x}}^2}
\ \Bigr )}^{\pm 1},
\end{equation}
the $(d,+1)$ and $(p,-1)$ subscripts and signs correspond to the deformation
potential and piezoelectric coupling respectively.

The self energy for the two-dimensional tunneling configuration is given by

\begin{equation}
\label{eq:d5}
\Sigma_{2}(p)
= \frac{\alpha_{d,p}}{(2 \pi)^3} \int_{-\infty}^{\infty}
dq_{z} {\vert M_{q_{z}} \vert}^2
\int_{-\infty}^{\infty} dq_{x}\int_{-\infty}^{\infty} dq_{y}
\frac{1}{\kappa(p-\imath q_{x})+
\frac{1}{2}q_{y}^2 }
{\Bigl (\sqrt{{q_{x}}^2 +{q_{y}}^2+{q_{z}}^2}\ \Bigr )}^{\pm 1},
\end{equation}
where the subscript {\scriptsize 2} will be used to define the two-dimensional case
throughout the paper.

\section{Deformation Potential Coupling}
\label{sec:deformation}

The self-energy integrals for the three-dimensional tunneling configuration
(\ref{eq:d4}) have an ultra-violet divergence for both coupling mechanisms.
Actually these divergences are cut off by the inverse lattice constant $a$,
but it goes against physical intuition when for a phenomena involving momenta much
smaller than the inverse lattice constant such cut-off is necessary. The
divergences can be eliminated with the help of the standard renormalization
procedure (see, e.g., Ref.\onlinecite{Itzykson80}). For the deformation 
potential coupling the large constant term
$\Sigma(0)$ (which has a form $\alpha_{d}(c_{0}/a^2)$ where $c_{0}$ is a
numerical constant) can be viewed as a renormalization of the potential
barrier height, $V$ and, hence, the renormalization of the propagation
constant $\kappa$,
\begin{equation}
\label{eq:e1}
\kappa_{ren}^2 = 2 m (V-E) + \Sigma_{d}(0).
\end{equation}
The abbreviation $ren$ is used to denote
renormalized quantities, also it is assumed that $\kappa$ is $\kappa_{ren}$
(unless otherwise stated), through the rest of the paper.
Actually the renormalization (\ref{eq:e1}) has to be
made before one passes to the parabolic approximation.

Canceling the non-$p$ dependent diverging part of $\Sigma(p)$ for the case of
deformation potential coupling does not cancel all the divergences. One needs
further to cancel a term  $\tilde{\Sigma_{d}}(0) p$, ($\tilde{\Sigma_{d}}(0)$
is logarithmically divergent, it can be estimated as $\alpha_{d}\kappa c_{1}
\ln(\kappa a)$ where $c_{1}$ is a numerical constant) which can be viewed as
a further renormalization due to a renormalization of the electron's
effective mass,

\begin{equation}
\label{eq:e2}
{\kappa \over m_{ren}}= {\kappa - \tilde{\Sigma}_{d}(0) \over m} \ .
\end{equation}
Subtracting the two diverging terms from the self-energy integral one obtains

\begin{equation}
\label{eq:e3}
\Sigma_{d}^{ren}(p)=
\Sigma_{d}(p)-\Sigma_{d}(0)- \tilde{\Sigma}_{d}(0)p =
-\alpha_{d}\frac{{\kappa}^2 {\lambda}^3}{6 {\pi}^2} \ln{2 \lambda} \ ,
\end{equation}
where $\lambda=p/\kappa$.

For the two-dimensional tunneling in the deformation potential coupling
the self-energy also has an ultra-violet divergence.
After subtracting the divergent term from the self-energy integral,
we obtain a finite expression, for the self-energy

\begin{equation}
\label{eq:e4}
\Sigma_{d,\small{2}}^{ren}(p)=
\Sigma_{d,\small{2}}(p)-\Sigma_{d,\small{2}}(0)=
-\alpha_{d} \frac{\sqrt{3/2}-1}{2 {\pi}^2} p \int_{-\infty}^{\infty}
dq_{z} {\vert M_{q_{z}} \vert}^2,
\end{equation}
For details of this calculation one is referred to the
Appendix \ref {ap:countor}.

In the tunneling problem, contrary to the propagation one, in the case of
weak coupling the Green function can be expanded in the self-energy. Then
after the substitution of Eqs. (\ref{eq:e3}) and (\ref{eq:e4}) into Eq.
(\ref{eq:gfm.pa.5}) the inverse Laplace transformation for the
three-dimensional case one gets that $t_{d}=1-(\alpha_{d}/6{\pi}^2)({\kappa}
L)^{-2}$. Thus the transition probability for the electron coupled via the
deformation potential coupling mechanism is given by

\begin{equation}
\label{eq:e5}
T= T^{0} \left[1-\frac{\alpha_{d}}{3{\pi}^2} \frac{1}{{(\kappa L)}^2}\right],
\end{equation}
where $T^{0}$ is the transition probability for the tunneling problem
without phonons.

In the two-dimensional tunneling case the transition probability is
given by

\begin{equation}
\label{eq:e6}
T_{\small 2}= T_{\small 2}^{0}
        \left[
    1-\alpha_{d}\frac{\sqrt{3/2}-1}{\pi^2 \kappa} \int_{-\infty}^{\infty}
    dq_{z} {\vert M_{q_{z}} \vert}^2
        \right] .
\end{equation}

\section{Piezoelectric Coupling}
\label{sec:piezoelectric}

In the two-dimensional case the integral for the piezoelectric self-energy is
finite. The finite independent of $p$ part of the integral, $\Sigma_{p}(0)
\sim (\alpha_{p}/d\kappa) {\vert M_{0} \vert}^2
\ln(d\kappa)$, can be viewed as
a finite renormalization of $\kappa$, where $M_{0}$ is the matrix element of
the phonon exponent, for $q_{z}=0$,
and $d$ is the length of the potential barrier in the $z$ direction. $1/d$ is
the scale that cuts off the $q_{z}$ integration, and it is assumed that in
the physical situation we are studying, $ 1 \ll d \kappa $.
Renormalizing $\Sigma$ we have

\begin{equation}
\label{eq:f1}
\Sigma_{p,\small{2}}^{ren}(p) =\Sigma_{p,\small{2}}(p)-\Sigma_{p,\small{2}}(0)=
\frac{ \alpha_{p}}{(2 \pi)^2}\ {p\over \kappa} \ {\vert M_{0} \vert}^2
\ln {(p d)}
\end{equation}

In the three-dimensional case the renormalization correction diverges in the
parabolic approximation and the renormalized self-energy is
\begin{equation}
\label{eq:f2}
\Sigma_{p}^{ren}(p)=
\Sigma_{p}(p)-\Sigma_{p}(0)=
\frac{\alpha_{p}}{2 {\pi}^2}\ \lambda \
\left (\ln{\lambda + \ln 2 - 1 } \right ).
\end{equation}
Details of the calculation of the integrals can be found Appendix
\ref{ap:countor}.

The terms $\lambda \ln \lambda$ in Eq.(\ref{eq:f2}) and $p\ln(pd)$ in
Eq.(\ref{eq:f1}) become large compared to $\kappa p$ as $p$ goes to zero.
Thus the perturbation theory in this case is not sufficient.

A similar problem was encountered by \'{E}del'shte\u{i}n when he studied the
piezoelectric polaron\cite{edelshtein73}. He considered high order vertex
corrections and showed that in each order of the perturbation theory the
highest power of the logarithm comes from diagrams with non-crossing phonon
lines. Neglecting all other diagrams he succeeded in obtaining an equation
for a vertex correction. Because the origin of the logarithmic divergence in
\'{E}del'shte\u{i}n's case and ours is the same we can use the same
approximation. In the following calculation we follow closely the derivation
presented in Ref.\onlinecite{edelshtein73}.

In the following discussion and calculation we will treat the three
dimensional case the treatment of the two dimensional case is very
similar therefore only the results shall be presented for it.

We start by considering vertex corrections.
The first vertex correction, $\Gamma_{1}$, presented in Fig. \ref{fig2}:

\begin{equation}
\label{eq:f3}
\Gamma_{1}(p_{1},{\bf k}_{1\perp},p_{2},{\bf k}_{2\perp}) =
\alpha_{p} \int \frac{d^3{\bf q}}{(2 \pi)^3}{1 \over q} \
    G_{{\bf k}_{1\perp}-{\bf q_{\perp}}}^{(0)}(p_{1}-\imath q_{x})
    G_{{\bf k}_{2\perp}-{\bf q_{\perp}}}^{(0)}
(p_{2}-\imath q_{x}).
\end{equation}
The integral is calculated in Appendix \ref{ap:leading-order} giving
\begin{equation}
\label{eq:f4}
\Gamma_{1} \approx - {\left (\alpha_{p}\over 2 {\pi}^2 {\kappa}^2\right )}
\ln{\left (\eta \over \kappa^2 \right )} \ .
\end{equation}
where $\eta=\kappa p+{k}^2_{\perp}/2$ \ ($p$ is either $p_{1}$ or $p_{2}$,
since under logarithmic accuracy the difference between them can be ignored,
for the same reason $k_{\perp}$ is either $k_{1\perp}$ or $k_{2\perp}$).
For $k_{\perp}=0$ the same result can
be obtained from the self-energy (\ref{eq:f2}) with the help of the Ward
identity.

For the two dimensional case the first vertex correction is
\begin{equation}
\label{eq:f5}
\Gamma_{1} \approx - {\left (\alpha_{p}\over 2 {\pi}^2 {\kappa}^2\right )}
{\vert M_{0} \vert}^2
\ln{\left ({ \tilde{\eta} \ d \over \kappa}\right )} \ ,
\end{equation}
where $\tilde{\eta}=\kappa p+{k}^2_{y}/2$. The calculation of this integral
can also be found in Appendix \ref{ap:leading-order}.

The second order diagrams for the vertex corrections are presented in
Fig.\ref{fig3}. The diagram (3,a) is reduced to the renormalization of the
Green function. It is possible to show that the leading order contribution to
the diagram (3,d) is of the order of
$(\alpha_{p}/2 \pi^{2}\kappa^{2})^{2}\ln(\eta/\kappa^2)$ containing lower
power of the logarithm than the power of $\alpha_{p}$. The leading order
contribution to the other two diagram is of the order of
$(\alpha_{p}/2 \pi^{2}\kappa^{2})^{2}\ln^{2}(\eta/\kappa^2)$, i.e., the power
of the logarithm equals the power of the coupling constant. The same relation
holds true for more complicated diagrams. That is, diagrams containing
intersections of phonon lines contain lower power of the logarithm than the
power of the coupling constant and diagrams without such an intersection get
another power of the logarithm in every next order of the perturbation
theory.

The last statement can be proven in the following way. The increase of the
order of the perturbation theory by unity in a vertex diagram without
intersection of phonon lines corresponds to the replacement of one of the
bare vertices in this diagram by $\Gamma_{1}$. Similar to the calculation in
Appendix \ref{ap:leading-order}, the values of the momentum in the contour of
this new $\Gamma_{1}$ that give the main contribution are much larger than
the momentum in external contours but much smaller than $\kappa$. So this
$\Gamma_{1}$ again gives a logarithmic contribution.

Another result of such a calculation is that in diagrams without phonon line
crossings the variables $p$ and $k_{\perp}$ enter only in the combination
$\eta=\kappa p+{k}^2_{\perp}/2$.

Thus when $\eta/\kappa^{2}$ is small enough, e.g. the effective potential
barrier is high enough, one can neglect diagrams containing phonon
intersection lines. The summation of all other diagrams leads to the equation
for the vertex $\Gamma$ presented in Fig.\ref{fig4}.

Since the Green function and the vertex correction are functions of $\eta$,
the Ward identity is expressed in the following form:
\begin{equation}
\label{eq:f6}
\frac{\partial {G}^{-1}}{\partial \eta}= \Gamma,
\end{equation}
Integrating Eq. (\ref{eq:f6}) and making use of the fact $\Gamma$ is a slow
function we get
\begin{equation}
\label{eq:f7}
G(\eta)= G^{(0)}(\eta) \ \Gamma^{-1}(\eta) \ .
\end{equation}
The substitution of this expression in the equation presented in
Fig.\ref{fig4} leads to the cancellation of two vertices and resulting in a
linear equation for the vertex function. The main contribution to the
integral term in this equation comes from the region where the momentum in
the contour in Fig.\ref{fig4} is much larger than the external momenta.
Similar to the calculation of $\Gamma_1$, the difference between the external
momenta is not important and they have to be kept only to limit the value of
otherwise logarithmically divergent integral. So the equation can be written
in the following form
\begin{equation}
\label{eq:f8}
\Gamma(\eta) = 1 +
    \alpha_{p} \int_{0}^{\infty} \frac{q dq}{(2 \pi)^2} \int_{-1}^{1} dt
    \frac{\Gamma(-\imath \kappa q t +q^2/2 +\eta ) }
    {{\left (-\imath \kappa q t + {q^2 \over 2} (1-t^2) +\eta
    \right ) }^2} \ .
\end{equation}
In this integral $t\sim1$ and the terms $q^{2}/2$ and $\eta$ are small
compared to $\kappa q$. So these terms can be neglected in the argument of
$\Gamma$ (it is a slow function) and $t^{2}$ can be neglected in the
coefficient of $q^{2}/2$ in the denominator (the exact value of this term is
not important). Now it is convenient to consider $\Gamma(\eta)$ as a function
of the large logarithmic variables $x=\ln(\eta/\kappa^2)$. Then keeping just
the large logarithm $x^{\prime}=\ln(q/\kappa)$ in the argument of $\Gamma$ in
the integrand we can easily integrate with respect to $t$. As a result, with
the logarithmic accuracy,
\begin{equation}
\label{eq:f9}
\Gamma(x)=1 - {\alpha_{p}\over 2 \pi^2 \kappa^2}
\int_{0}^{x} dx' \Gamma(x'),
\end{equation}
The solution of this equation and using Eq.(\ref{eq:f9}) gives

\begin{equation}\displaystyle
\label{eq:f10}
G(\eta)= {1 \over \eta}
    {\left(\eta \over {\kappa}^2\right)}^{\alpha_{p}\over 2\pi^2 \kappa^2}.
\end{equation}

In the two dimensional case the Green function is given by:
\begin{equation}\displaystyle
\label{eq:f11}
G(\tilde{\eta})= {1 \over \tilde{\eta}}
    {\left(\tilde{\eta} \ d \over \kappa \right)}^{{\alpha_{p}\over 2\pi^2 \kappa^2}
{\vert M_{0} \vert}^2}.
\end{equation}

Inserting these results in the equation for the transition amplitude
(\ref{eq:gfm.pa.5}) and performing the inverse Laplace transform
\cite{bateman53} we obtain for the three-dimensional tunneling configuration

\begin{equation}
\label{eq:f12}
T= T^{0} \ (\kappa L)^{-{\alpha_{p}\over \pi^2 {\kappa}^2}} \ .
\end{equation}
The two-dimensional case gives the following transition probability
\begin{equation}
\label{eq:f13}
T= T^{0} \ {1 \over (\kappa L)}
{\left ( {d \over L}\right )}^{-{\alpha_{p}\over \pi^2 {\kappa}^2}
{\vert M_{0} \vert}^2} \ .
\end{equation}

\section{Summary and Conclusions}
\label{sec:summary}

In this paper we have presented a detailed study of the effect of coupling of
an electron tunneling across a rectangular barrier, to the zero-point
fluctuations of the acoustic phonon field. Two coupling mechanisms, the
piezoelectric and deformation potential, were studied for two and three
dimensional tunneling configurations. We considered a small coupling and used
two main approximations: the static approximation and the parabolic
approximation.

Contrary to the one-dimensional case \cite{pazy99}, the main effect of the
deformation potential coupling is the barrier and the electron mass
renormalization. On the top of this the second order perturbation theory is
sufficient. As a result only a prefactor in the tunneling probability is
changes.

For the piezoelectric coupling mechanism, the situation is quite different.
The perturbation theory is not sufficient and an accurate calculation of the
vertex is necessary. As a result of the coupling, in both two-dimensional and
three-dimensional cases the tunneling probability acquires a prefactor
proportional to a power of the barrier width. The small coupling constants
enters in the power. This result resembles piezoelectric polaron. When the
barrier is wide enough the piezoelectric coupling can lead to a substantial
reduction of the tunneling probability.

It is interesting also to compare the sign of the coupling effect in one- and
higher dimensions. In one-dimensional case the coupling led to an enhancement
of the tunneling probability. The physics behind this is that an electron
chooses for tunneling the most favorable phonon configuration.\cite{pazy99}
In high-dimensional case, due to ultra-violet divergences, such favorable
phonon configurations led to overall renormalization of the barrier which
cannot be separated from the bare barrier. The measurable effect depending on
the length of the barrier is a reduction of the tunneling probability.

In both cases the effects mentioned are enhanced when tunneling is near the
top of the barrier, though as the effective potential barrier height, $V-E$,
is lowered two effects become important: dynamical corrections to our static
approximation and an effect which can be described as channeling, the change of the
electron wave function due to the interaction with phonons. These are
subjects for future study.

\section{Acknowledgments}

The work was supported by The Israel Science Foundation, grant No. 174/98.

\newpage

\appendix\section
{calculation of the self-energies}
\label{ap:countor}

In this appendix we present the main stages in the calculation of the
self-energy integrals appearing in the text, except the self-energy integral
for the three-dimensional deformation potential coupling which is straight
forwardly evaluated. We start with the calculation of the self-energy
integral, for the deformation potential coupling in the two-dimensional case,
after the subtraction of the divergent term.
\begin {eqnarray}
\label{eq:ap.1a}
\Sigma_{2}^{ren}(p) & = &
\frac{\alpha_{d}}{{(2 \pi)}^3}\
\int_{-\infty}^{\infty}dq_{z} {\vert M_{q_{z}} \vert}^2
\int_{-\infty}^{\infty} dq_{x}
\int_{-\infty}^{\infty}
d q_{y} \left ( \frac{1}{\kappa p -\imath \kappa q_{x} + \frac{1}{2}
q_{y}^2} - \frac{1}{-\imath \kappa q_{x} + \frac{1}{2}
q_{y}^2} \right ) \nonumber \\
&\times &\sqrt{q_{x}^2 + q_{y}^2+q_{z}^2} \ .
\end{eqnarray}
Noticing that for the $q_{x}$ integration poles are located in the lower half
of the complex $q_{x}$ plane we shift the $q_{x}$ integration contour to the
upper half plane. The shifted contour goes from $i\infty$ along the imaginary
axis, rounds the point $x_{0}=i\sqrt{q_{y}^2+q_{z}^2}$ in the positive
direction and comes back to $i\infty$. Thus we obtain
\begin {eqnarray}
\label{eq:ap.2a}
\Sigma_{2}^{ren} (p) & = & \frac{\alpha_{d}}{4 {\pi}^3}\
\int_{-\infty}^{\infty}dq_{z} {\vert M_{q_{z}} \vert}^2
\int_{-\infty}^{\infty} dq_{y}  \nonumber \\
&\times & \int_{\sqrt{q_{y}^2+q_{z}^2}}^{\kappa}
dx \left ( \frac{1}{\kappa p + \kappa x + \frac{1}{2}
q_{y}^2} - \frac{1}{\kappa x + \frac{1}{2}
q_{y}^2} \right )  \sqrt{x^2 - q_{y}^2-q_{z}^2} \ .
\end{eqnarray}
Where the upper cut-off, $\kappa$, for the $x$ integration is due to the
parabolic approximation. The integration with respect to $q_{z}$ is cut off
by the matrix element at $q_{z}\sim1/d$ where $d$ is the size of the
quantization in the $z$ direction. We assume that $d\kappa\gg 1$ and hence
the main contribution to the integral (\ref{eq:ap.2a}) comes from the region
$q_{y}\gg q_{z}$. Then we can neglect the $q_{z}^2$ appearing in the square
roots, and
\begin {eqnarray}
\label{eq:ap.3a}
\Sigma_{2}^{ren}(p) & = &
-\frac{\alpha_{d}\kappa p}{4 {\pi}^3}\
\int_{-\infty}^{\infty}dq_{z} {\vert M_{q_{z}} \vert}^2
\int_{-\infty}^{\infty} dq_{y}
\int_{\vert q_{y} \vert}^{\kappa}
dx \frac{\sqrt{x^2 - q_{y}^2}}{(\kappa p + \kappa x + \frac{1}{2}
q_{y}^2) (\kappa x + \frac{1}{2}
q_{y}^2)}   \ .
\end{eqnarray}
Since we are interested in the limit $p\sim1/L$ which is much smaller than
all other scales, we neglect $\kappa p$ denominator, scaling out $\kappa$ by
changing variables $x$ to $\kappa x$ and $q_{y}$ to $\kappa y$ we arrive at
\begin {eqnarray}
\label{eq:ap.4a}
\Sigma_{2}^{ren}(p) & = &
-\frac{\alpha_{d}  p}{2 {\pi}^3}\
\int_{-\infty}^{\infty}dq_{z} {\vert M_{q_{z}} \vert}^2
\int_{0}^{\infty} dy
\int_{\vert y \vert}^{1}
dx \frac{\sqrt{x^2 - y^2}}{ (x + \frac{1}{2}
y^2)^2}   \ .
\end{eqnarray}
Changing variables to: $x=t y$ and changing the order of integration
with respect to $t$ and $y$ we get
\begin {eqnarray}
\label{eq:ap.5a}
\Sigma_{2}^{ren}(p) & = &
-\frac{\alpha_{d}  p}{2 {\pi}^3}\
\int_{-\infty}^{\infty}dq_{z} {\vert M_{q_{z}} \vert}^2
\int_{1}^{\infty}dt \sqrt{t^2 -1}
\int_{0}^{1/t}
\frac{dy}{ (t + \frac{1}{2} y)^2}   \ .
\end{eqnarray}
Were this integral is already easily evaluated resulting in Eq. (\ref{eq:e4}).

We now evaluate the self energy term for the piezoelectric coupling in the
two-dimensional case,
\begin {eqnarray}
\label{eq:ap.6a}
\Sigma_{2}^{ren}(p) & = &
\frac{\alpha_{p}}{{(2 \pi)}^3}\
\int_{-\infty}^{\infty}dq_{z} {\vert M_{q_{z}} \vert}^2
\int_{-\infty}^{\infty} dq_{x}
\int_{-\infty}^{\infty}
d q_{y} \left ( \frac{1}{\kappa p -\imath \kappa q_{x} + \frac{1}{2}
q_{y}^2} - \frac{1}{-\imath \kappa q_{x} + \frac{1}{2}
q_{y}^2} \right )  \nonumber \\
&\times & {1\over \sqrt{q_{x}^2 + q_{y}^2+q_{z}^2}} \ .
\end{eqnarray}
Using the same contour integration we derive
\begin {eqnarray}
\label{eq:ap.7a}
\Sigma_{2}^{ren}(p) & = &
-\frac{\alpha_{p}\kappa p}{2 {\pi}^3}\
\int_{-\infty}^{\infty}dq_{z} {\vert M_{q_{z}} \vert}^2
\int_{0}^{\infty} dq_{y}
\int_{\sqrt{q_{y}^2+q_{z}^2}}^{\infty}
 \frac{dx }{(\kappa p + \kappa x + \frac{1}{2}
q_{y}^2) (\kappa x + \frac{1}{2}
q_{y}^2)} \nonumber \\
&\times &   {1\over \sqrt{x^2 - q_{y}^2-q_{z}^2}} \ .
\end{eqnarray}
$q_{y}$ in the integrand is smaller than $x$ and as we will see $x\ll\kappa$.
Due to this  we neglect $q_{y}^2$ compared to $\kappa x$ and we obtain
\begin {eqnarray}
\label{eq:ap.8a}
\Sigma_{2}^{ren}(p) & = &
-\frac{\alpha_{p} p}{2  {\pi}^3 \kappa}\
\int_{-\infty}^{\infty}dq_{z} {\vert M_{q_{z}} \vert}^2
\int_{0}^{\infty} dq_{y}
\int_{\sqrt{q_{y}^2+q_{z}^2}}^{\infty}
 \frac{dx }{ \sqrt{x^2 - q_{y}^2-q_{z}^2}}\ {1\over (p + x)x}  \ .
\end{eqnarray}
Changing variables to: $x=t \sqrt{q_{y}^2+q_{z}^2} \ ; \ q_{y}= y q_{z}$, we get
\begin {eqnarray}
\label{eq:ap.9a}
\Sigma_{2}^{ren}(p) & = &
-\frac{\alpha_{p} p}{2  {\pi}^3 \kappa}\
\int_{-\infty}^{\infty}dq_{z} {\vert M_{q_{z}} \vert}^2
\int_{0}^{\infty} \frac{d{y}}{\sqrt{y^2+1}}
\int_{1}^{\infty}
 \frac{dt}{t\sqrt{t^2-1} }\ {1\over p + t\sqrt{y^2+1}q_{z} } \ .
\end{eqnarray}
When $p\ll1/d$ the integral with respect to $q_{z}$ contains a large
logarithm. In the logarithmic approximation
\begin {eqnarray}
\label{eq:ap.9b}
\Sigma_{2}^{ren}(p) & = &
-\frac{\alpha_{p} p}{\pi^3 \kappa} {\vert M_{0} \vert}^2
\int_{p}^{1/d}{dq_{z} \over q_{z}}
\int_{0}^{\infty} \frac{d{y}}{y^2+1}
\int_{1}^{\infty}
 \frac{dt}{t^{2}\sqrt{t^2-1} } \ ,
\end{eqnarray}
which results in Eq. (\ref{eq:f1}).

Finally we calculate the self-energy for piezoelectric coupling in the
three-dimensional case,
\begin {eqnarray}
\label{eq:ap.10a}
\Sigma_{p}^{ren}(p) & = &
\frac{\alpha_{p}}{{(2 \pi)}^3}\ \int_{-\infty}^{\infty} dq_{x}
\int d^2 q_{\perp} \left ( \frac{1}{\kappa p -\imath \kappa q_{x} + \frac{1}{2}
q_{\perp}^2} - \frac{1}{-\imath \kappa q_{x} + \frac{1}{2}
q_{\perp}^2} \right ) {1 \over \sqrt{q_{x}^2 + q_{\perp}^2}}
\end{eqnarray}
Performing the angle integration and scaling out $\kappa$ and changing
variables to: $q_{\perp}^2=2\kappa pz ; \ q_{x}= px$, we get

\begin{equation}
\label{eq:ap.11a}
\Sigma_{p}^{ren}(p)=
-\frac{\alpha_{p}}{{(2 \pi)}^2}\
\int_{0}^{\infty} dz \int_{-\infty}^{\infty} dx
    \frac{1}{(1-\imath x +z)(- \imath x +z)}
    {1 \over \sqrt{x^2 + 2z/\lambda}} ,
\end{equation}
where $\lambda=p/\kappa$. Noticing that for the $x$ integration poles are
located in the lower half of the complex $x$ plane we shift the contour as we
did before. Now the shifted contour rounds the point $x=i\sqrt{2z/\lambda}$
and the result becomes
\begin{eqnarray}
\label{eq:ap.12a}
\Sigma_{p}^{ren}(p) & = & -\frac{\alpha_{p}}{2 {\pi}^2}\
    \int_{0}^{\infty} dz \int_{0}^{\infty} dy
    \frac{1}{\left (1+\sqrt{2z\over \lambda}+z+ y \right )
    \left (\sqrt{2z\over \lambda}+z +y \right )}
    {1 \over \sqrt{2y\sqrt{2z\over \lambda}+y^2}}
\nonumber \\ & = &
    -\frac{\alpha_{p}\lambda}{2\pi^2}
    \int_{0}^{\infty} \frac{dt}{\sqrt{(2+t)t}}
    \int_{0}^{\infty}
    \frac{du}{\left(1+ u + {\lambda u^2\over 2}+ ut\right)
    \left(1 + {\lambda u\over 2} + t\right)}.
\end{eqnarray}
In the integral with respect to $u$ we neglect terms of the order of
$\lambda$ so that
\begin{eqnarray}
\label{eq:ap.a13}
\Sigma_{p}^{ren}(p) & = &
    -\frac{\alpha_{p}\lambda}{2\pi^2}
    \int_{0}^{\infty} \frac{dt}{\sqrt{(2+t)t}} \ {1 \over (1 + t)^{2}}
    \left[\ln{2\over\lambda} - 1 + 2 \ln(1 + t)\right]
\nonumber \\ & = &
    -\frac{\alpha_{p}\lambda}{2\pi^2}
    \left[\ln{2\over\lambda} - 1 + 2 (1 - \ln 2)\right] .
\end{eqnarray}

\section
{Leading order calculation of first vertex corrections}
\label{ap:leading-order}

Under logarithmic accuracy the main contribution to the integral
(\ref{eq:f3}) comes from the region $\kappa\gg q\gg p,{\bf k}_{\perp}$. For
this reason difference between $p_1$ and $p_2$ as well as between ${\bf
k}_{1\perp}$ and ${\bf k}_{2\perp}$ is not important. So the integral
(\ref{eq:f3}) can be written as
\begin{equation}
\label{eq:ap.1b}
\Gamma_{1}(\eta) =
    \alpha_{p} \int_{0}^{\infty} \frac{q dq}{(2 \pi)^2} \int_{0}^{\pi}
    \frac{d\theta \sin{\theta}}
    {{\left (-\imath q\kappa \cos{\theta} + {q^2 \over 2} \sin^2{\theta} +\eta
    \right ) }^2}\ \ ,
\end{equation}
where $\eta=\kappa
p+k_{\perp}^{2}/2$. The term $q^{2}/2$ cuts off the integration with
respect to $q$ from above and the term $\eta$ does the same from below. In
the region between these limits the integral is logarithmic and its value is
accumulated from the whole interval of the integration. So that the exact
value of these limits is not important and we can simplify the integration by
the replacement of $\sin^{2}\theta$ with unity. As a result
\begin{equation}
\label{eq:ap.2b}
\Gamma_{1}(\eta) = \frac{2 \alpha_{p}}{(2 \pi)^2 \kappa^2} \int_{0}^{\infty} \
\frac{q dq}{q^2 + {\left ( {q^2 \over 2\kappa}+{\eta \over \kappa} \right ) }^2}
    = - {\alpha_{p} \over 2 \pi^2 {\kappa}^2}
\ln {\left ( {\eta \over \kappa^2} \right )} .
\end{equation}

Where a lograthmic approximation has been used in calculating the last integral.

Considering the first vertex correction in the two dimensional case again
under logarithmic accuracy the difference between $p_1$ and $p_2$ as well
as between ${\bf k}_{1y}$ and ${\bf k}_{2y}$ is not important. Thus the first
vertex correction is given by
\begin{equation}
\label{eq:ap.3b}
\Gamma_{1}(\tilde{\eta}) =
    \frac{\alpha_{p}}{(2 \pi)^3}
    \int_{-\infty}^{\infty} {dq_{z}} {\vert M_{q_{z}} \vert}^2
    \int_{-\infty}^{\infty} {dq_{x}}
    \int_{-\infty}^{\infty} \frac{dq_{y}}{\sqrt{q_{x}^2+q_{y}^2+q_{z}^2}}
    \frac{1}
    {{\left (-\imath q_{x}\kappa  + {q_{y}^2 \over 2}  + \tilde{\eta}
    \right ) }^2}\ \ ,
\end{equation}
where $\tilde{\eta}=\kappa p+{k}^2_{y}/2$.
Using the contour integration described in Appendix (\ref{ap:countor}) one gets
an equation which is very similar to Eq. (\ref{eq:ap.7a})
\begin {equation}
\label{eq:ap.4b}
\Gamma_{1}(\tilde{\eta}) =
\frac{\alpha_{p}}{2 {\pi}^3}\
\int_{-\infty}^{\infty}dq_{z} {\vert M_{q_{z}} \vert}^2
\int_{0}^{\infty} dq_{y}
\int_{\sqrt{q_{y}^2+q_{z}^2}}^{\infty}
 \frac{dx }{ (\kappa x + \frac{1}{2}
q_{y}^2 +\tilde{\eta})^2 } \nonumber \\
{1\over \sqrt{x^2 - q_{y}^2-q_{z}^2}} \ .
\end{equation}
Treating this integral in the same way Eq. (\ref{eq:ap.7a}) has been
evaluated results in Eq. (\ref{eq:f5}). The main difference between this
$\Gamma_{1}$ and that for three dimensional case is that now the logarithmic
integral is cut off from above by the upper limit of $q_{z}$ which is of the
order of $1/d\ll\kappa$.

\newpage

\begin{figure}
\caption{The self energy term. The solid line represents the bare
electron propagator whereas the dashed line represents interaction with
phonons. }
\label{fig1}
\end{figure}

\begin{figure}
\caption{The first correction to the vertex $\Gamma $.}
\label{fig2}
\end{figure}

\begin{figure}
\caption{Second order vertex corrections}
\label{fig3}
\end{figure}

\begin{figure}
\caption{The equation for the renormalized vertex $\Gamma$ resulting from
the summation of leading logarithmic corrections. Double lines represent the
full electron Green function.}
\label{fig4}
\end{figure}


\begin{thebibliography}{99}

\bibitem{Caldeira 81} A.O. Caldeira and A.J. Leggett, Phys.\ Rev.\ Lett.\
{\bf 46}, 211 (1981); Phys.\ Rev.\ A {\bf 31}, 1059 (1985); Ann.\ Phys. (N.Y.)
{\bf 149}, 374 (1983); A.J. Leggett, Phys.\ Rev. \ B {\bf 30}, 1208 (1984).

\bibitem{Perrson88} B.N.J. Persson and A. Baratoff, Phys.\ Rev.\ B {\bf 38},
9616 (1988); D.V. Averin, {\em ibid.}\ {\bf 50}, 8934 (1994)
; M. Ueda and T. Ando,  Phys.\ Rev.\ Lett.\
{\bf 72}, 1726 (1994).

\bibitem{Grabert92}Yu.V. Nazarov,  Sov.\ Phys-JETP {\bf 68}, 561 (1989);
Yu.V. Nazarov, Phys.\ Rev.\ B {\bf 43}, 6220 (1991);
G.-L. Ingold and Yu.V. Nazarov, in {\em {Single Charge Tunneling }} \ edited by
H. Grabert and M. Devoret (Plenum, New York, 1992); M.H. Devoret, D. Esteve,
H. Grabert, G.-L. Ingold, H. Pothier and C. Urbina Phys.\ Rev.\ Lett.\ {\bf 64},
1824 (1990); S.M. Girvin, L.I. Glazman, M. Jonson, D.R. Penn and M.D. Stiles,
{\em ibid.}\ {\bf 64}, 3183 (1990).

\bibitem{Devoret92}M. H. Devoret, D. Esteve, C. Urbina, J. Martins, J. Martinis,
A. Cleland and J. Clarke, in {\em {Quantum Tunneling in Condensed Media}} \
edited by Yu. Kagan and A.J. Leggett (Elsevier, Amsterdam, 1992).

\bibitem{Marclay88} P. Gu\'{e}ret, E. Marclay and H. Meier, Solid State
Commun. {\bf 68}, 977 (1988); D.Esteve, J.M. Martinis, C. Urbina, E.
Turlot, and M.H. Devoret, Phys. Scr. {\bf T29}, 121 (1989).

\bibitem{Bak86} R. Bruinsma and P. Bak, Phys.\ Rev.\ Lett.\ {\bf 56},
420 (1986); B.Y. Gelfand, S. Schmitt-Rink, and a.F.J.Levi,
{\em ibid.}\ {\bf 62}, 1683 (1989);
M. Ueda, Phys.\ Rev.\ B {\bf 54}, 8676 (1996).

\bibitem{re:polaron}
For a review of the polaron problem, see T.K. Mitra, A. Chatterjee, and S.
Mukhopadhyay, Phys.\ Rep. {\bf 153} 91 (1987).

\bibitem{pazy99} E. Pazy and B. Laikhtman,  Phys.\ Rev.\ B {\bf 59},
15854 (1999).

\bibitem{re:isotropic}
A.R. Hutson, J. \ Appl. \ Phys. {\bf 32}, 2287 (1961).

\bibitem{Babic91} V.M. Babich, V.S. Buldyrev   {\em {Asymptotic Methods
in Shortwave Diffraction Theory}}\
({Springer-Verlag}, {New-York}, 1991).

\bibitem{Itzykson80} C.L. Itzykson, J. Zuber {\em {Quantum Field Theory}}\
({Mcgraw-Hill}, {New-York}, 1985).

\bibitem{edelshtein73} V.M. \'{E}del'shte\u{i}n, Sov.\ Phys-JETP {\bf 36},
809 (1973).

\bibitem{bateman53} H. Bateman, {\em {Higher Transcendental Functions}},\
Vol. 1 ({Robert E. Krieger},{Florida} 1981).

\end{thebibliography}
\end{document}